\documentclass[lettersize,journal]{IEEEtran}

\usepackage{amsmath,amsfonts}
\usepackage{array}
\usepackage{textcomp}

\usepackage{url}
\usepackage{verbatim}
\usepackage{booktabs}
\usepackage{graphicx}
\usepackage{subfigure}
\usepackage{stfloats}
\usepackage{cite}
\usepackage{bm}

\usepackage[ruled,linesnumbered,vlined]{algorithm2e}

\usepackage{hyperref}

\usepackage[rgb]{xcolor}

\begin{document}

\title{Deep Graph Reinforcement Learning for UAV-Enabled Multi-User Secure Communications}

\author{Xiao Tang, Kexin Zhao, Chao Shen, Qinghe Du, Yichen Wang, Dusit Niyato, and Zhu Han
\thanks{X. Tang is with the School of Information and Communication Engineering, Xi'an Jiaotong University, Xi'an 710049, China, and also with the Research \& Development Institute of Northwestern Polytechnical University in Shenzhen, Shenzhen 518063, China. (e-mail: tangxiao@xjtu.edu.cn)}
\thanks{K. Zhao is with the School of Electronics and Information, Northwestern Polytechinical University, Xi'an 710072, China.}
\thanks{C. Shen is with the Faculty of Electronics and Information Engineering, Xi’an Jiaotong University, Xi’an 710049, China.}
\thanks{Q. Du and Y. Chen are with the School of Information and Communication Engineering, Xi'an Jiaotong University, Xi'an 710049, China.}
\thanks{D. Niyato with the College of Computing and Data Science, Nanyang Technological University, Singapore.}
\thanks{Z. Han is with the Department of Electrical and Computer Engineering, University of Houston, Houston 77004, USA.}
}



\maketitle

\begin{abstract}
While unmanned aerial vehicles (UAVs) with flexible mobility are envisioned to enhance physical layer security in wireless communications, the efficient security design that adapts to such high network dynamics is rather challenging. The conventional approaches extended from optimization perspectives are usually quite involved, especially when jointly considering factors in different scales such as deployment and transmission in UAV-related scenarios. In this paper, we address the UAV-enabled multi-user secure communications by proposing a deep graph reinforcement learning framework. Specifically, we reinterpret the security beamforming as a graph neural network (GNN) learning task, where mutual interference among users is managed through the message-passing mechanism. Then, the UAV deployment is obtained through soft actor-critic reinforcement learning, where the GNN-based security beamforming is exploited to guide the deployment strategy update. Simulation results demonstrate that the proposed approach achieves near-optimal security performance and significantly enhances the efficiency of strategy determination. Moreover, the deep graph reinforcement learning framework offers a scalable solution, adaptable to various network scenarios and configurations, establishing a robust basis for information security in UAV-enabled communications.
\end{abstract}


\section{Introduction}

Unmanned aerial vehicles (UAVs) have become an increasingly important role in future 6G networks with their rapid development and wide applications~\cite{uav-com,s3}. The transformative capabilities in terms of coverage, flexibility, and adaptability envisioned through UAVs in wireless network necessitate a heightened need for information security~\cite{uav-sec}. In this regard, physical layer security that exploits channel characteristics for secrecy has emerged as a promising solution with low-complexity security implementation~\cite{pls,t1}. The keyless operations alleviate the burden for intensive computation and key management under conventional cryptographic methods, which is particularly fascinating for UAV-enabled communications with limited resources and high dynamics~\cite{uav-pls}.

Accordingly, the UAV-enabled communication with physical layer security has received wide attention with insightful results~\cite{uav-pls-survey,t3}. In this respect, the mainstream research efforts have been devoted to the optimization-based or heuristic approaches, which, though providing effective security provisioning, often suffer from high computational demand with multi-round iterations to approach some suboptimal solutions~\cite{uav-opt-survey}. This issue becomes remarkably challenging in UAV-enabled communications as the high dynamics of UAVs induce severely fluctuating channel conditions, where the optimization-based approaches may struggle to strive with prompt adaptation~\cite{s1,uav-unct-2}. Therefore, there is an urgent need for efficient physical layer security design for UAV-enabled communications to facilitate the applications across civilian and military domains.

Meanwhile, deep learning has found increasingly wide applications in wireless area recently to address the computational and scalability challenges faced by traditional methods in secure communication design~\cite{dl-sec-survey,s4}. Deep learning models excel in capturing complex relationship and patterns, allowing for data-driven approaches for security and enabling offline training with online inference. These features are rather attractive to fit the UAV-enabled systems, where the environmental conditions and network topologies change rapidly and learning-based models can be trained to optimize the performance efficiently~\cite{uav-nn-survey}. Specifically, for UAV-enabled multi-user secure communications, the legitimate users and eavesdroppers affect each other and the mutual interference acts the pivotal factor to be tackled. This is when graph neural networks (GNNs), as a special type of deep learning model established on graph-structure data, become particularly pertinent~\cite{uav-gnn-1,uav-gnn-2}. In this context, different roles in the network along with the inherent interference are interpreted as graphs, and the message-passing mechanisms in GNNs enable to track their mutual influence and evolution. By leveraging GNN-based learning, scalable and adaptive approaches can be developed to efficiently secure the UAV-enabled networks.

Furthermore, different from infrastructure-based low-mobility communications, the flexible deployment of UAVs introduces significant dynamics. Consequently, the UAV location induced large-scale channel attenuation and the small-scale fading need to be jointly considered to achieve effective security design. In this respect, deep reinforcement learning provides an effective approach to dynamically position the UAVs, achieving the refinement of the large-scale factor to maximize the security gain~\cite{rl-gnn-1,t2}. Unlike deep learning model, the reinforcement learning interacts with the environment to update the deployment, which accommodates the highly variable scenarios involving UAVs~\cite{rl-gnn-2,s2}. Therefore, with combined graph and reinforcement learning techniques, we can exploit the graph structure to track the interfering user interactions, and employ the reinforcement mechanism to adapt to different network scenarios, which facilitates the effective strategy design to meet real-time security needs in dynamic network environments.

In this paper, we consider the UAV-enabled multi-user communications with physical layer security provisioning. Different from existing works, we adopt a hierarchical learning model to jointly investigate the UAV deployment and transmission beamforming to maximize the secrecy rate. The main contributions are summarized as follows:
\begin{itemize}
	\item We consider the UAV-enabled multi-user secure communications with a joint design of deployment and transmission strategies. We then propose a layered decomposition to facilitate the analysis that, the outer layer seeks for deployment as the large-scale factor and inner layer solves beamforming as the small-scale factor, which are further combined to provide the security solution.
	\item We reinterpret the considered communication network as a graph structure, and establish the GNN model for security beamforming. The message passing mechanism is established to track the interference relationships among the users for efficient beamforming output. Also, the proposed GNN model features permutation equivalence property to facilitate training and generalization.
	\item The outer-layer UAV deployment problem is addressed using the soft actor-critic (SAC) reinforcement learning approach. The learning agent exploits the GNN-based beamforming to evaluate the rewards and interacts with the environment to achieve security-oriented deployment in changing network topologies.
	\item Extensive numerical results validate the effectiveness of our proposal, demonstrating the near-optimal secrecy performance across various scenarios. Also, the results highlight the scalability and adaptability to cover new scenarios and the computational efficiency as compared with the baselines.
\end{itemize}

The rest of this paper is organized as follows. In Sec.~\ref{sec:rw}, we review the related work. Sec.~\ref{sec:sys} introduces the UAV-enabled multi-user system model with secure communication problem formulation. Sec.~\ref{sec:gnn} discusses the GNN-based secure beamforming and Sec.~\ref{sec:drl} proposes the deep reinforcement learning-based UAV deployment. The numerical results are provided in Sec.~\ref{sec:sim}, and finally this paper is concluded in Sec.~\ref{sec:con}.

\section{Related Work} \label{sec:rw}

Information security is one of the fundamental concerns for UAV-enabled communications, for which the physical layer security technique featuring keyless operations has raised wide interest. Intuitively, we can employ the classical designs of physical layer security, such as security beamforming~\cite{uav-sec-bf}, friendly relaying~\cite{uav-sec-relay}, artificial jamming~\cite{uav-sec-an}, statistical security~\cite{uav-sec-sta}, in the context of UAV scenarios to protect the transmissions. Different from conventional-infrastructure communications, the high mobility of UAVs provides a new dimension for secrecy and has been actively exploited to combat eavesdropping. In~\cite{uav-sec-deploy-1}, the authors propose a UAV-aided space-air-ground communications network for information secrecy, demonstrating significant security enhancement through joint UAV deployment and resource optimization. In~\cite{uav-sec-deploy-2}, the authors investigate the strategic deployment of UAV swarm as aerial base stations, and optimize UAV positioning, user association, and power allocation to maximize secrecy performance. In~\cite{uav-sec-deploy-3}, the authors address the security issue for UAV edge computing, and solve energy-efficient secure offloading problems against the eavesdroppers. In~\cite{uav-sec-traj-1}, the authors investigate secure transmissions of aerial underlay Internet of Things systems, where the average secrecy rate is maximized by optimizing UAV trajectory and resources while considering eavesdropping uncertainty. In~\cite{uav-sec-traj-2}, the authors propose a robust design for UAV trajectory and artificial jamming power optimization to maximize the minimum average secrecy rate. In~\cite{uav-sec-traj-3}, the authors propose a secure short-packet communication system with a UAV-enabled mobile relay through joint design of coding blocklengths, transmit powers, and UAV trajectory to maximize secrecy throughput. More recently, the UAV secure communications are also jointly explored with emerging techniques for more flexible and robust designs. In~\cite{uav-sec-game}, the authors establish the game model for the anti-eavesdropping interactions in UAV communications, where the pursuit-evasion process is undertaken towards the equilibrium to evaluate the security performance. In~\cite{uav-sec-ris}, the authors exploit reconfigurable intelligent surfaces for secured UAV communications with joint optimization of beamforming, jamming and reflection to downgrade the eavesdropping. In these studies, the UAV dynamics are jointly considered with the resources in different dimensions for security-oriented designs. Consequently, the joint investigation of factors in different aspects in UAV-enabled communications generally induces complicated optimization problems. It usually requires multi-round iterations and approximations to approach the solution, which can be operationally intricate and difficult to adapt in dynamic UAV communication scenarios.

Meanwhile, learning-based security solutions have become increasingly popular in UAV-enabled communications, with the applications of various learning model, where the deep neural networks can be employed to approximate the optimal decision makings. In~\cite{uav-dl-1}, the authors present a framework combining optimization and deep learning for UAV-assisted wireless-powered secure communications, where the output of the trained neural network serves as the initial values with optimization-based fine-tune. In~\cite{uav-dl-3}, the authors develop a deep learning-driven 3D robust beamforming method for secure UAV communication, achieving improved beam steering under partial channel conditions. In~\cite{uav-dl-2}, the authors investigate the secure UAV-relaying network, the deep neural networks are exploited to assist the deployment and transmission strategies. Besides the classical deep model, the GNNs with explicit graph structure can more effectively track the mutually interacting behavior in wireless systems. In~\cite{uav-gnn-2}, the authors propose a GNN-based approach for sum-rate maximization in multi-user network, a complex residual graph attention network is established to achieve effective beamforming with fast response time and scalability. In~\cite{uav-gnn-1}, the authors introduce a GNN architecture for joint multicast and unicast beamforming, improving the system performance and scaling well to different network settings. In~\cite{uav-gnn-3}, the authors study the reflection-assisted communications and leverage a GNN architecture to construct the mapping from pilot sequences to the transmission and reflection strategies to maximize the system rate. Moreover, the deep reinforcement learning technique enables interactions with the environment and updates the policy based on the obtained reward, which allows more flexible manners to learn to adapt in UAV-enabled communications. In~\cite{uav-rl-1}, the authors consider UAV-relay communications, where GNN-enabled link selection and reinforcement learning-based trajectory are jointly exploited to maximize the number of active users. In~\cite{uav-rl-2}, the authors propose a GNN-empowered multi-agent reinforcement learning to manage the age of information in multi-UAV networks, with distributed UAV trajectory optimization in unknown environments. In~\cite{uav-rl-3}, the authors propose a graph-attention-based reinforcement learning framework for UAV trajectory design and resource assignment, achieving improved convergence and optimal cumulative rewards. Based on the analysis above, we can see that the learning-based designs for UAV-enabled communications are emerging rapidly, while the security concerns remain under-explored. As GNN and reinforcement learning models are capable of representing the mutual interference and adversarial interactions underlying secure communications involving dynamic UAVs, it motivates our endeavor to leverage these techniques to achieve effective information secrecy for UAV-enabled communications.

\section{System Model} \label{sec:sys}

\begin{figure}[t]
   \centering
   \includegraphics[width=0.45\textwidth]{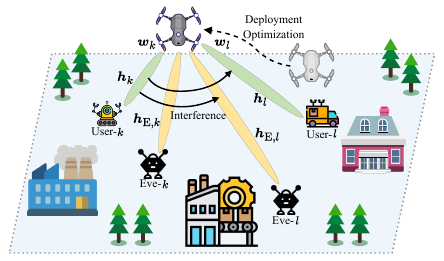} 
   \caption{System model.}
   \label{fig:sys}
\end{figure}

We consider a UAV-enabled multi-user downlink communication system within an area denoted by $ \Lambda $, where a single UAV serves as a mobile base station, transmitting confidential message to $K$ legitimate users, denoted by $ \mathcal{K} = \left\{1,2,\ldots, K\right\} $, as shown in Fig.~\ref{fig:sys}. Meanwhile the legitimate communications are subject to eavesdropping threats, for which we assume each legitimate receiver is associated with one specific wiretapper and we abuse the notation $ \mathcal{K} $ without ambiguity. The UAV as aerial base station is equipped with $ N $ antennas, and the legitimate users and eavesdroppers are of single antenna, which establishes the multi-user multi-input single-output secure communications. This setup has been well established as the studies in~\cite{1to1-1,1to1-2}, which can be justified by associating the geographically closest adversary as the corresponding eavesdropper or adopting the strong eavesdropping assumption. Also, as shown later, our approach can be conveniently extended to other eavesdropping models. The UAV operates within the area with horizontal coordinates denoted by $ \bm{q}_U = \left( x_{\text{U}}, y_{\text{U}} \right) \in \Lambda $, while maintaining a fixed altitude $H$. Meanwhile, the legitimate user-$k$ is located with horizontal coordinates $ \bm{q}_k = \left( x_k, y_k \right) \in \Lambda$, and the corresponding eavesdropper is positioned at $ \bm{q}_{\text{E}, k} =  \left( x_{\text{E}, k}, y_{\text{E}, k} \right)\in \Lambda$. In this paper, we assume fixed ground node locations while the UAV is enabled with flexible deployment for security enhancement. The channel condition between the UAV and legitimate user-$k$ is denoted as $ \bm{h}_k \left( \bm{q}_U \right) $, and similarly, the channel to the corresponding eavesdropper is $ \bm{h}_{\text{E}, k} \left( \bm{q}_U \right) $, as a function of UAV location in the network.

For the UAV-enabled communications, a data signal $s_k$ is transmitted to legitimate user-$k$ with $\mathbb{E}[|s_k|^2] = 1$, by adopting a dedicated beamforming vector $\bm{w}_k \in \mathbb{C}^{N}$. Then, the transmitted signal from the UAV, denoted by $\bm{x}$, is represented as
\begin{equation}
\bm{x} = \sum_{k\in\mathcal{K}} \bm{w}_k^{\dag} s_k,
\end{equation}
where transmit beamformer is subject to the power constraint that
\begin{equation}
\sum\limits_{k\in\mathcal{K}} \|\bm{w}_k\|^2 \leq P^{\max},
\end{equation}
with $P^{\max}$ being the maximum allowed power at the UAV. Accordingly, the legitimate user-$k$ receives a signal $y_k$ that includes both the intended signal and interference from signals directed to other users, given as,
\begin{equation}
y_k = \bm{h}_{k}^{\dag} \bm{w}_k s_k + \bm{h}_{k}^{\dag}\sum_{l\in\mathcal{K}\backslash\{k\}}  \bm{w}_l s_l + n_k,
\end{equation}
where the last term is the Gaussian noise with $n_k \sim \mathcal{CN}(0, \sigma^2_0)$. The received signal at eavesdropper-$k$ is similarly obtained by replacing the channel $ \left\{\bm{h}_k\right\}_{k\in\mathcal{K}} $ with $ \left\{\bm{h}_{\text{E},k}\right\}_{k\in\mathcal{K}} $, where the first term is the intended wiretap signal, and the rest corresponds to the experienced interference and noise.

Based on the transmission model, the legitimate transmission of user-$k$ has an achievable communication rate $R_k$ as
\begin{equation}
R_k = \log \left( 1 + \frac{|\bm{h}_{k}^{\dag} \bm{w}_k|^2}{\sum\limits_{l\in\mathcal{K}\backslash \{k\}} |\bm{h}_{k}^{\dag} \bm{w}_l|^2 + \sigma^2_0} \right).
\end{equation}
Similarly, the wiretap rate at the corresponding eavesdropper is obtained as
\begin{equation}
R_{\text{E}, k} = \log \left( 1 + \frac{|\bm{h}_{\text{E}, k}^{\dag} \bm{w}_k|^2}{\sum\limits_{l\in\mathcal{K}\backslash \{k\}} |\bm{h}_{\text{E}, k}^{\dag} \bm{w}_l|^2 + \sigma^2_0} \right).
\end{equation}
Thus, the secrecy rate for legitimate user-$k$ is derived as
\begin{equation} \label{eq:sec-rate}
R_k^{\text{sec}} = \left( R_k - R_{\text{E}, k} \right)^+,
\end{equation}
where $(\cdot)^+ = \max(\cdot, 0)$ ensures non-negative secrecy rate.

The goal of this work is to maximize the total secrecy rate across all legitimate users by jointly optimizing the beamforming vectors and the UAV deployment, which leads to the problem formulation as
\begin{IEEEeqnarray} {cl}
\max_{\{\bm{w}_k\}_{k\in\mathcal{K}}, \bm{q}_U } & \quad  \sum\limits_{k\in\mathcal{K}} R_k^{\text{sec}}\\
\rm{s.t.} & \quad \sum\limits_{k\in\mathcal{K}} \|\bm{w}_k\|^2 \leq P^{\max}, \\
& \quad \bm{q}_U \in\Lambda.
\end{IEEEeqnarray}

Although the formulated problem appears straightforward, it is non-convex and thus challenging to solve in an efficient manner, particularly when considering the UAV dynamics and environment adaptation. First, the joint optimization problem involves both the UAV deployment and the beamforming vectors, interdependent in a complicated manner within the non-convex problem. The existing proposals are mostly based on optimization that generally requires multi-round relaxations and approximations and can be quite time-consuming, especially when the dimension of the problem becomes large. Second, the UAV-enabled communications are inherently dynamic, for which traditional optimization methods often require re-solving the problem each time the environment changes. As a result, optimization-based approaches may fail to adapt quickly and thus can be unsuitable for the cases requiring prompt security decision-makings. Third, the exiting optimization-based methods typically depend on problem-specific formulations and assumptions, which limits their scalability to varying network conditions or different numbers of users and eavesdroppers. The lack of generalization hinders their practical deployment in dynamic and diverse UAV-enabled networks.

Therefore, in this work, we propose a learning-based framework to find the physical layer security solution for UAV-enabled multi-user communications. As the considered issues of deployment and beamforming are taken as the large-scale and small-scale factors in the network, respectively, we decompose the problem in a double-layer structure to tackle different factors. Specifically, the inner layer addresses the security beamforming considering fixed UAV deployment with GNN-based interpretation and training, and the outer layer determines the UAV deployment by incorporating the GNN-based beamforming under a deep reinforcement learning framework. In this respect, we exploit the GNN to approximate the complex relationships between users, eavesdroppers, and the UAV in a non-convex optimization setting, and interpret the secure beamforming problem as a graph-based learning task. Moreover, the integration of DRL allows the UAV to learn optimal positioning strategies directly through interactions with the environment, adapting to changes in network environments in real time.

\section{GNN-Based Security Beamforming} \label{sec:gnn}

In this section, we address the inner problem as security beamforming through GNN-based learning, on condition of fixed UAV deployment at the outer layer. Accordingly, the problem is specified as
\begin{IEEEeqnarray} {cl}
\max_{\{\bm{w}_k\}_{k\in\mathcal{K}}} & \quad \sum\limits_{k\in\mathcal{K}} R_k^{\text{sec}}\\
\rm{s.t.} & \quad \sum\limits_{k\in\mathcal{K}} \|\bm{w}_k\|^2 \leq P^{\max}. \label{eq:p-max}
\end{IEEEeqnarray}
Evidently, there are complex interference among the legitimate users and eavesdroppers, which needs to be tackled properly so as to maximize the secrecy rate for each legitimate user while minimizing the information leakage to the eavesdroppers. From a learning perspective, the intended neural network is fed with the channel condition as input, and produce the corresponding security beamforming design, given by $ \bm{W} = \Phi\left( \bm{H}; \bm{\Theta} \right) $, where $ \bm{W} $, $ \bm{H} $, and $ \bm{\Theta} $ denote the collective beamforming, channel condition, and neural network parameters, respectively. In this respect, the network parameters are trained so as to maximize the sum secrecy rate as $ R^{\text{sec}} (\Phi\left( \bm{H}; \bm{\Theta} \right), \bm{H}) $, where $ R^{\text{sec}} = \sum\nolimits_{k\in\mathcal{K}} R_k^{\text{sec}} $. To this end, we adopt the graph model to capture the non-Euclidean structure of physical network, and establish the GNN-based learning mechanism to obtain the security beamformer.

\subsection{GNN Architecture}

To leverage GNNs for secure beamforming design, we represent the elements the UAV-enabled communication network as graph components. As we can see, the spatial and interference relationships between legitimate users, eavesdroppers, and the UAV, can be reflected through the channel condition in the network with obtained secrecy rates at the legitimate users. Accordingly, we established a graph with $ K $ nodes, and each node-$k\in\mathcal{K}$ is associated with a feature as
\begin{equation}
\begin{aligned}
\bm{z}_k = \left[ \mathfrak{Re}\left(\bm{h}_k\right)^{\top}, \mathfrak{Im}\left(\bm{h}_k\right)^{\top}, \mathfrak{Re}\left(\bm{h}_{\text{E},k}\right)^{\top}, \mathfrak{Im}\left(\bm{h}_{\text{E},k}\right)^{\top} \right. \\
\left. \mathfrak{Re}\left(\bm{e}_k\right)^{\top}, \mathfrak{Im}\left(\bm{e}_k\right)^{\top}\right]^{\top},
\end{aligned}
\end{equation}
by concatenating the channel conditions of legitimate user-$k$ and its eavesdropper, along with an additional embedding $ \bm{e}_k \in \mathbb{C}^{N} $, with $ \mathfrak{Re}\left(\cdot\right) $ and $ \mathfrak{Im}\left(\cdot\right) $ denoting the real and imaginary parts of complex variables, respectively. Since the network channel condition is incorporated within the collective node features, we establish the graph without explicitly considering the edge model. Here, the node feature is constructed by the locally ``useful'' information (useful for the legitimate transmission, and useful for the corresponding eavesdropper), and the interference is obtained through the further interactions among nodes. Also, the nodes are of equivalent model, which facilitates the desired permutation equivalence property during GNN-based learning as elaborated below.

\begin{figure*}[t]
   \centering
   \includegraphics[width=0.95\textwidth]{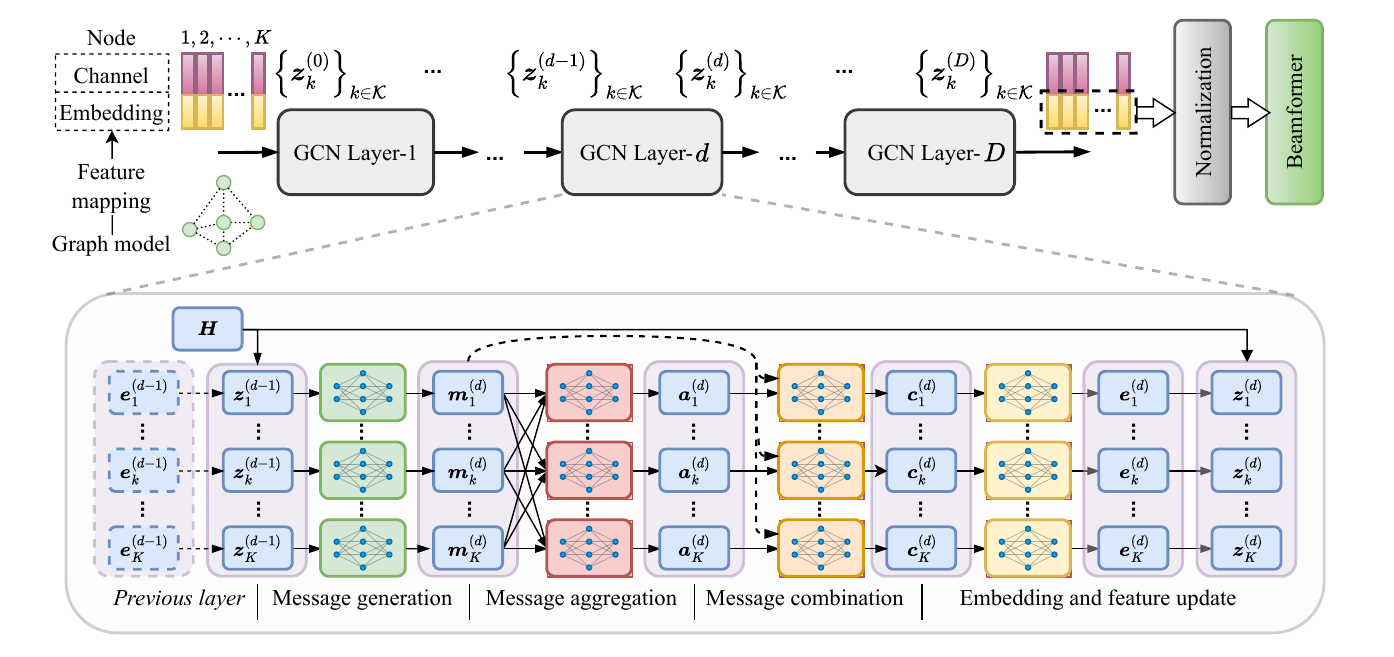} 
   \caption{GNN architecture for security beamforming.}
   \label{fig:gnn}
\end{figure*}

With graph-based representation of the physical network, the proposed GNN architecture is constructed with $ D $ cascaded graph convolutional network (GCN) layers along with a normalization layer, as shown in Fig.~\ref{fig:gnn}. In each GCN layer, the consecutive operations of message generation, aggregation, and combination are conducted to implement the message-passing mechanism of GNN. In accordance with the definition of node feature, during the layered learning process, only the embedding is updated while the channel information is preserved. This is motivated by the idea to achieve improved representation of network condition during learning by preserving the channel information. Therefore, for the $d$-th GCN layer, the input is denoted by $ \left\{ \bm{z}_k^{(d-1)} \right\}_{k\in\mathcal{K}} $ with
\begin{equation} \label{eq:mesg-gen}
\begin{aligned}
\bm{z}_k^{(d-1)} = \left[ \mathfrak{Re}\left(\bm{h}_k\right)^{\top}, \mathfrak{Im}\left(\bm{h}_k\right)^{\top}, \mathfrak{Re}\left(\bm{h}_{\text{E},k}\right)^{\top}, \mathfrak{Im}\left(\bm{h}_{\text{E},k}\right)^{\top} \right. \\
\left. \mathfrak{Re}\left(\bm{e}_k^{(d-1)}\right)^{\top}, \mathfrak{Im}\left(\bm{e}_k^{(d-1)}\right)^{\top}\right]^{\top},
\end{aligned}
\end{equation}
where $ \bm{e}_k^{(d-1)} $ is the obtained embedding of node-$k$ in the previous layer. Then, the input is fed into a linear layer for message generation, given as
\begin{equation}
\bm{m}_k^{(d)} = f_{\text{gen},k}^{(d)} \left( \bm{z}_k^{(d)}; \bm{\Theta}_{\text{gen},k}^{(d)} \right)
\end{equation}
where $\bm{m}_k^{(d)}$ denotes the generated message through the operation $ f_{\text{gen},k}^{(d)} $, implemented as a fully-connected layer with parameter $ \bm{\Theta}_{\text{gen},k}^{(d)} $.

Then, the generated messages at each node are shared with the others through aggregation and combination operations, given in a general form as
\begin{equation}
\begin{aligned}
\bm{c}_k^{(d)} = f_{\text{comb},k}^{(d)} \left( \bm{m}_k^{(d)}, f_{\text{aggr},k}^{(d)} \left( \left\{ \bm{m}_l^{(d)} \right\}_{l\in\mathcal{N}(k)}; \right.\right. \qquad \\  \left.\left. \bm{\Theta}_{\text{aggr},k}^{(d)} \right); 
 \bm{\Theta}_{\text{comb},k}^{(d)} \right),	
\end{aligned}
\end{equation}
where $ \bm{c}_k^{(d)} $ is the achived message representation after aggregation and combination functions, denoted by $ f_{\text{aggr},k}^{(d)} $ and $ f_{\text{comb},k}^{(d)} $ with parameters $\bm{\Theta}_{\text{aggr},k}^{(d)} $ and $ \bm{\Theta}_{\text{comb},k}^{(d)} $, respectively, 
with $ \mathcal{N}(k) $ being the set of neighbors of node-$k$. To ensure the scalability and generalization of proposed GNN architecture, it is critical to choose appropriate implementations of aggregation and combination~\cite{uav-gnn-3,gnn-mark-1}. In this respect, the message aggregation at node-$k$, is conducted according to
\begin{equation}
\bm{a}_k^{(d)} = f_{\text{aggr},k}^{(d)} \left( \left\{ \bm{m}_l^{(d)} \right\}_{l\in\mathcal{K}\backslash\{k\}}; \bm{\Theta}_{\text{aggr},k}^{(d)} \right),
\end{equation}
where $\bm{a}_k^{(d)}$ denotes the obtained aggregation while incorporating all other nodes as the neighbors, without loss of generality. The messages from the neighbors are processed through a neural network with parameters $ \bm{\Theta}_{\text{aggr},k}^{(d)} $ and fed into the aggregation function to get $ \bm{a}_k^{(d)} $. The aggregation function usually takes the form such as $ \mathsf{max} $, $ \mathsf{sum} $, or $ \mathsf{mean} $, to ensure permutation-invariant property with respect to the inputs. Here, we adopt the maximization operation due to the fact that the aggregate interference is usually dominated by the heaviest interferer. Then, the nodes combines its own message with the aggregation from others as
\begin{equation}
	\bm{c}_k^{(d)} = f_{\text{comb},k}^{(d)} \left( \bm{m}_k^{(d)}, \bm{a}_k^{(d)}; \bm{\Theta}_{\text{aggr},k}^{(d)} \right),
\end{equation}
to reach a combined message $ \bm{c}_k^{(d)} $. Here we simply concatenate the messages to preserve the representation learned in previous operations. This update process refines the node representation by integrating information from its neighbors, gradually enhancing its ability to determine optimal beamforming in response to the network interference pattern.

As the output of current layer, the combined message at each node is fed into a multilayer perceptron with output fitting the dimension of the node embedding such that
\begin{equation}
	\left[\mathfrak{Re}\left(\bm{e}_k^{(d)}\right)^{\top}, \mathfrak{Im}\left(\bm{e}_k^{(d)}\right)^{\top}\right]^{\top} = f_{\text{out},k}^{(d)} \left( \bm{c}_k^{(d)}; \bm{\Theta}_{\text{out},k}^{(d)} \right).
\end{equation}
where $ f_{\text{out},k}^{(d)} $ denotes the output function with network parameters $\bm{\Theta}_{\text{out},k}^{(d)}$. The obtained embedding $ \bm{e}_k^{(d)} $ is concatenated with the channel condition to constructed the output of current layer, i.e., the input of the subsequent GCN layer, in the similar manner as~(\ref{eq:mesg-gen}).

For the message generation, aggregation, and combination operations introduced above as implementation of message-passing mechanism of GNN, we can see that only the embedding of nodes are updated. In this regard, we keep the channel state information in the learning processes and thus preserve its direct influence over the update of embedding. As such, the characteristics of the network environment can be sufficiently represented in the node embedding. Meanwhile, as the embedding appears in complex-valued form (which is seen later to directly lead to the beamformer), we adopt the parametric rectified linear unit ($\mathsf{PReLU}$) as the activation function in the neural network layers. This activation function has both positive and negative output, fitting the mathematical requirement of the embedding. Also, the learnable weights within $\mathsf{PReLU}$ improve adaptation to channel coefficient data, which may vary in a wide range since the users in a large area is connected to one single UAV. Moreover, the message-passing mechanism is repeated across multiple GCN layers, enabling each node to progressively aggregate information from the distant nodes. After $D$ layers of message passing, the node features contain a rich representation of the network environment and interference relationships, which can be used to generate the beamforming vectors.

Finally, the obtained embedding of each node at the last layer, i.e., $ \left\{ \bm{e}_k^{(D)} \right\}_{k\in\mathcal{K}} $, is exploited to obtain the beamforming vector. As the beamforming is required to satisfy the power constraint in~(\ref{eq:p-max}), the normalization operation is conducted such that
\begin{equation} \label{eq:w-out}
	\bm{w}_k = P^{\max} \frac{\mathfrak{Re}(\bm{e}_k^{(D)}) + \jmath \mathfrak{Im}(\bm{e}_k^{(D)}) }{\sqrt{ \sum\limits_{k\in\mathcal{K}} \left\| \bm{e}_k^{(D)} \right\|^2 } }   ,
\end{equation}
where $ \jmath $ is the imaginary unit, and $ \bm{e}_k^{(D)} $ is the final-layer node embedding output. The obtained beamforming can then be fed into the loss function for the GNN training.

\subsection{Loss Function and Training}

The GNN model presented above takes the reinterpreted network condition as input to produce the security beamforming so as to maximize the secrecy rate. Accordingly, we define the loss function based on the sum secrecy rate overall users as
\begin{equation}
	\mathsf{L}\left( \bm{\Theta} \right) = -\mathbb{E}_{\bm{H}\sim\mathcal{H}} R^{\text{sec}} (\Phi\left( \bm{H}; \bm{\Theta} \right), \bm{H}),
\end{equation}
where the GNN-based mapping $ \Phi $ is established with the collected neural network parameters $ \bm{\Theta} = \left\{ \bm{\Theta}_{\text{gen},k}^{(d)}, \bm{\Theta}_{\text{aggr},k}^{(d)} , \bm{\Theta}_{\text{comb},k}^{(d)} , \bm{\Theta}_{\text{out},k}^{(d)} \right\}_{\forall d, \forall k} $, and $ \mathcal{H} $ is the conforming distribution of the collective channel condition over the network.

As we can see above, to facilitate the GNN implementation with respect to complex-valued channel condition and beamformer, we separate the real and imaginary parts such that the GNN can be calculated in real domain. Technically, we can reconstruct the complex-valued beamformer as~(\ref{eq:w-out}), and obtain the secrecy rate as~(\ref{eq:sec-rate}). However, in this manner, the gradients need to be calculated with respect to complex-valued variables, which incurs additional difficulties for neural network training. To this issue, we separate the real and imaginary parts of the complex-valued components in secrecy rate and introduce
\begin{equation} \label{eq:gamma}
	\gamma_{kl} = \begin{bmatrix}
		\mathfrak{Re}(\bm{h}_k^{\dag}) & -\mathfrak{Im}(\bm{h}_k^{\dag}) \\
		\mathfrak{Im}(\bm{h}_k^{\dag}) & \mathfrak{Re}(\bm{h}_k^{\dag})
	\end{bmatrix}
	\begin{bmatrix}
	\mathfrak{Re}(\bm{w}_k) \\ \mathfrak{Im}(\bm{w}_k)
	\end{bmatrix},
\end{equation}
and similarly for $ \gamma_{\text{E},kl} $ by replacing $ \bm{h}_k $ with $ \bm{h}_{\text{E},k} $. Then, the secrecy rate can be recast as
\begin{equation}
\begin{aligned}
R_k^{\text{sec}} = & \left( \log\left( 1 + \frac{\left\| \gamma_{kk} \right\|^2}{\sum\limits_{l\in\mathcal{K}\backslash\{k\}} \left\| \gamma_{kl} \right\|^2 + \sigma_0^2}  \right) \right. \\
& \quad \left. - \log\left( 1 + \frac{\left\| \gamma_{\text{E},kk} \right\|^2}{\sum\limits_{l\in\mathcal{K}\backslash\{k\}} \left\| \gamma_{\text{E},kl} \right\|^2 + \sigma_0^2}  \right) \right)^+,
\end{aligned}
\end{equation}
which can be calculated along with~(\ref{eq:gamma}) for real-domain operations.

The GNN model is trained in an unsupervised learning framework, where each training sample includes network instances with specific user and eavesdropper distributions, UAV position, and thus channel condition. During each training iteration, the following steps are performed. First, for each network instance, the UAV-enabled network is mapped to a graph representation with node features defined before. Second, the GNN performs message passing across the graph, updating node features through multiple GCN layers to produce the final node representations, which are then mapped to the security beamformer. Third, the secrecy rate-based loss is calculated based on the generated beamforming vectors and current network channel condition. Last, the gradients of the loss function with respect to the GNN parameters are computed using back propagation method. Here, we adopt the optimizer through stochastic gradient descent, minimizing the loss and improving model performance. Based on existing studies~\cite{sgd}, with standard assumptions on the loss function, i.e., Lipschitz gradients and bounded variance, the stochastic gradient descent converges to a stationary point with a proper learning rate. While in our model, the non-convex loss induces fluctuating gradient updates in different regions, and the graph-based massage passing introduces correlations among nodes, making it intricate for a rigorous convergence proof. However, due to the effectiveness of stochastic gradient descent approach with over-parameterization mechanism in the proposed GNN, the learning process is rather likely to converge to some high-quality local optima, as evidenced in the numerical results.

Moreover, we can conveniently verify that the proposed GNN architecture features permutation equivalence, as the message generation, aggregation, combination, and output are all conducted in permutation-equivalent manners. This property enables reuse of training samples by simply shifting the indices and thus improves the training efficiency. More importantly, the GNN with permutation equivalence property can be conveniently generalized to cover various network settings and adapts to different distributions of users and eavesdroppers.

As a further remark, the proposed GNN architecture for security beamforming can be easily extended to the different network scenarios exampled below. First, while this work focuses on the one-to-one correspondence between legitimate users and eavesdroppers, we may further consider the general multi-eavesdropper or colluding eavesdropping models. In this regard, we can preserve current GNN architecture and adopt the corresponding worst-case secrecy rate when calculating the loss function, which is further back propagated to train the neural network. Second, when the receivers are also equipped with multiple antennas, which constitutes multiple-input-multiple-output transmissions, we can introduce the receiver beamformer in the node embedding, and update the neural network accordingly to cover the more complicated transceiving cases. Finally, we may also further consider the integration of friendly jammer or reconfigurable intelligent surface. In this regard, we may introduce an additional neural network module which connects with all the existing nodes, mimicking the interactions with the jammer/reconfigurable intelligent surface to calculate the secrecy rate, and train the corresponding neural network in similar manners. These extensions highlight the flexibility of the GNN-based framework, making it adaptable to a wide range of realistic or complicated network scenarios.

\section{SAC-Based UAV Deployment} \label{sec:drl}

\begin{figure*}[t]
   \centering
   \includegraphics[width=0.95\textwidth]{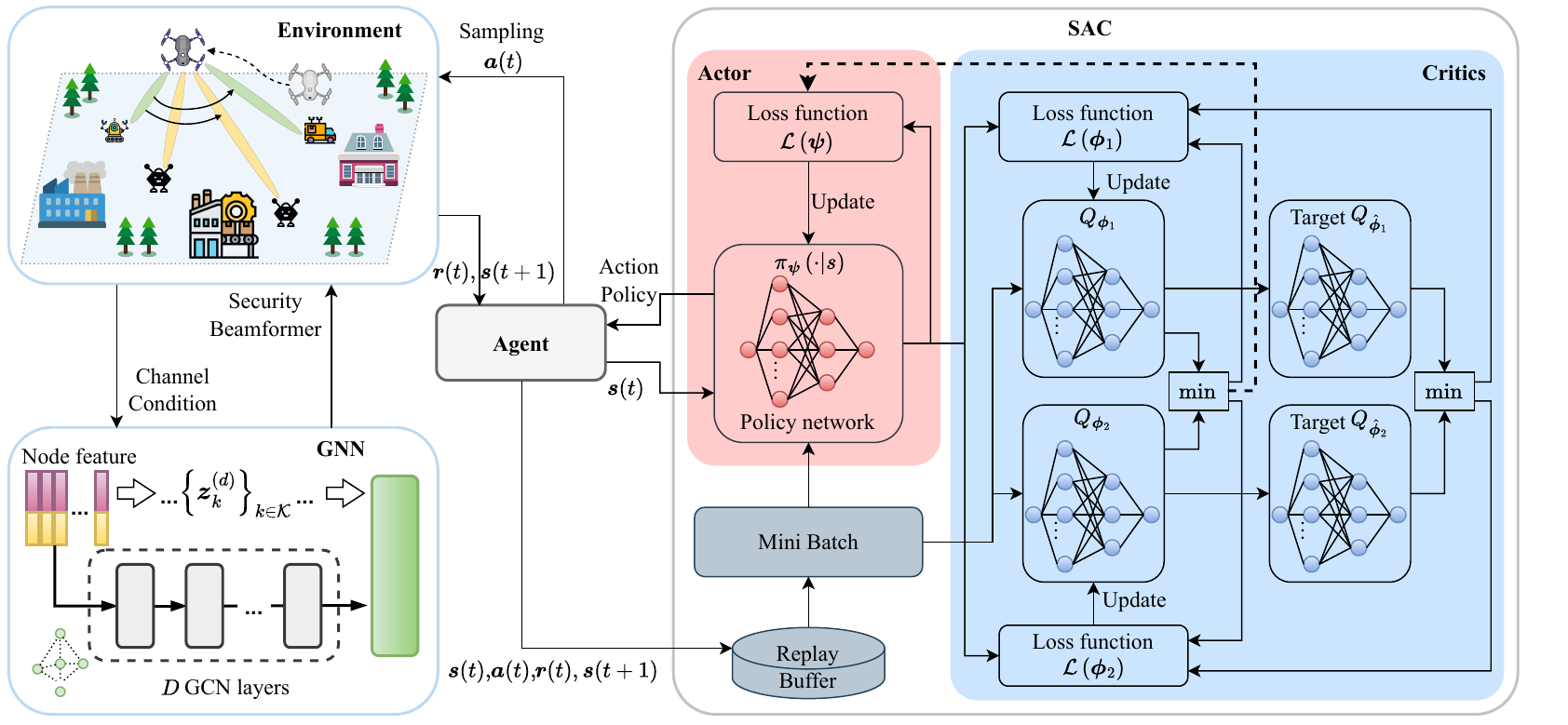} 
   \caption{SAC architecture for UAV deployment.}
   \label{fig:sac}
\end{figure*}

Based on the problem decomposition presented before, in this section, we investigate the outer-layer UAV deployment while incorporating the inner-layer GNN-based security beamforming. Accordingly, the problem is specified as
\begin{IEEEeqnarray} {cl}
\max_{\bm{q}_U} & \quad  \sum\limits_{k\in\mathcal{K}} R_k^{\text{sec}} \left( \left\{ \bm{h}_{ k} \left( \bm{q}_U \right) \right\}_{k\in\mathcal{K}}, \left\{ \bm{h}_{\text{E}, k} \left( \bm{q}_U \right) \right\}_{k\in\mathcal{K}} \right)\\
\rm{s.t.} & \quad \bm{W} = \Phi\left( \bm{H}\left( \bm{q}_U \right); \bm{\Theta} \right), \label{eq:gnn-con} \\
& \quad \bm{q}_U \in\Lambda,
\end{IEEEeqnarray}
where the constraint in~(\ref{eq:gnn-con}) indicates the GNN-based beamforming, with explicit UAV deployment-based channel condition $ \bm{H}\left( \bm{q}_U \right) $ to obtain the beamforming $ \bm{W} $.
Generally, the UAV deployment significantly impacts the security performance by influencing large-scale channel characteristics and spatial interference patterns. As such, the outer problem poses unique challenges due to its dynamic nature and high dimensionality, making traditional optimization approaches hardly feasible for real-time applications. Towards this issue, we adopt the soft actor-critic (SAC) reinforcement learning algorithm, which handles continuous state-action spaces and encourages efficient exploration through entropy regularization. By integrating SAC with the GNN-based inner-layer beamforming, the UAV deployment adapts dynamically to user positions and improves the system secrecy rate.

\subsection{UAV Deployment as a Markov Decision Process}

To tackle the UAV deployment issue for secrecy rate maximization, we first recast the problem as a Markov decision process (MDP). The MDP is established over a time series denoted by $ \mathcal{T} = \left\{ 1, 2, \cdots, t, \cdots, T \right\} $, where the UAV acts as the agent seeking for deployment optimization. Meanwhile, MDP model compromises the following components:

\subsubsection{State Space}
The state space $\mathcal{S}$ captures current UAV deployment configuration and the network environment. At time step $t$, the state $\bm{s}(t)$ is represented as
\begin{equation}
\bm{s}(t) = \left\{ (x_{\text{U}}, y_{\text{U}}), \{\bm{h}_k\}_{k\in\mathcal{K}}, \{\bm{h}_{\text{E},k}\}_{k\in\mathcal{K}} \right\},
\end{equation}
where the variables are also specified with argument of time-$t$ yet omitted for notation simplicity.

\subsubsection{Action Space}
The action space $\mathcal{A}$ defines the set of possible movements for the UAV. At each time step, the UAV takes an action $\bm{a}(t)$ to adjust its position as
\begin{equation}
\bm{a}(t) = (\Delta x, \Delta y),
\end{equation}
where $\Delta x$ and $\Delta y$ are the changes in the UAV horizontal coordinates along $x$- and $y$-axes, respectively. The action space is continuous, allowing the UAV to reach a fine-tuned deployment. At a certain time instance-$t$, the action $\bm{a}(t)$ induces an updated deployment as
\begin{equation}
(x_{\text{U}}(t+1), y_{\text{U}}(t+1)) = (x_{\text{U}}(t) + \Delta x, y_{\text{U}}(t) + \Delta y).
\end{equation}

\subsubsection{Reward Function}
The reward function evaluates the quality of the UAV deployment in terms of the security performance. Accordingly, we adopt the system secrecy rate as the instantaneous reward, i.e., $ r(\bm{s}(t), \bm{a}(t)) = R^{\text{sec}} $, where current deployment and the GNN-based security beamformer are jointly exploited for reward calculation. Then, the UAV agent learns a policy $\pi(\bm{a}_t | \bm{s}_t)$ that maximizes the expected cumulative reward as
\begin{equation}
J(\pi) = \mathbb{E}_{\pi} \left[ \sum_{t\in\mathcal{T}} \eta^t {r}(\bm{s}(t), \bm{a}(t)) \right],
\end{equation}
where $\eta \in [0, 1]$ is the discount factor that balances immediate and long-term rewards.

\subsection{Soft Actor-Critic Learning Architecture}

To solve the UAV deployment in the form of a MDP, we employ the SAC-based learning approach. Compared with conventional deep reinforcement learning, SAC is designed for continuous state-action spaces, where its entropy-regularized objective encourages policy exploration, leading to more robust and stable learning processes.

\subsubsection{Overview of SAC Architecture}
The overall architecture of the proposed SAC framework is shown in Fig.~\ref{fig:sac}, which consists of three main components explained below. First, the actor network: this network with parameter $ \bm{\psi} $ approximates the policy $\pi_{\bm{\psi}}(\bm{a}(t)|\bm{s}(t))$, which outputs a probabilistic distribution over actions for a given state. As the policy is stochastic, it enables effective exploration of the action space. Second, the critic networks: there are two Q-value approximators, $Q_{\bm{\phi}_1}(\bm{s}(t), \bm{a}(t))$ and $Q_{\bm{\phi}_2}(\bm{s}(t), \bm{a}(t))$, to estimate the expected cumulative reward of a state-action pair, with parameters $ \bm{\phi}_1 $ and $ \bm{\phi}_2 $, respectively. Meanwhile, there are two counterparts as the target critic networks with parameters $ \hat{\bm{\phi}}_1 $ and $ \hat{\bm{\phi}}_2 $, respectively, for target Q-value as $Q_{\hat{\bm{\phi}}_1}(\bm{s}(t), \bm{a}(t))$ and $Q_{\hat{\bm{\phi}}_2}(\bm{s}(t), \bm{a}(t))$. The double-critic structure helps mitigate potential overestimation issue while training. Third, the entropy regularization: an entropy term is introduced into the policy objective to encourage exploration, given as
\begin{equation}
	\mathcal{E}(\pi(\cdot|\bm{s}(t)) = -\mathbb{E}_{\bm{a}(t)\sim\pi(\cdot|\bm{s}(t))} \left[ \log\left( \pi(\bm{a}(t)|\bm{s}(t)) \right) \right],
\end{equation}
which is calculated to promote diverse action selection for reward maximization. Accordingly, the introduced entropy term leads to the revised value functions as
\begin{equation}
	V (\bm{s}) = \mathbb{E}_{\bm{a}\sim\pi(\cdot|\bm{s})} \left[ Q\left(\bm{s}, \bm{a}  \right) - \alpha \log \left(\pi(\bm{a}|\bm{s}\right) \right],
\end{equation}
for an initial state $ \bm{s} $ along with a temperature parameter $\alpha$ as the weight of the entropy term, and the Q-function is defined with respect to an action $ \bm{a} $ as
\begin{equation}
	Q(\bm{s},\bm{a}) = r(\bm{s},\bm{a}) + \eta \mathbb{E}_{\bm{s}^\prime \sim \mathcal{P}(\cdot | \bm{s}, \bm{a})} \left[V(\bm{s}^\prime)\right],
\end{equation}
with $ \mathcal{P}(\cdot | \bm{s}, \bm{a}) $ being the state transition probability due to the taken action in current state.


\subsubsection{Loss Functions}
In accordance with components incorporated within the SAC learning architecture, the loss functions are specified as below. First, for the entropy term, the temperature parameter is tuned while learning to minimize the loss as
\begin{equation} \label{eq:loss-entr}
   \mathcal{L}(\alpha) = \mathbb{E}_{{\bm{a}(t)\sim\pi(\cdot|\bm{s}(t))}} \left[ -\alpha \log \pi(\bm{a}(t)|\bm{s}(t)) - \bar{\mathcal{E}} \right],
\end{equation}
where $\bar{\mathcal{E}}$ is the target entropy that determines the degree of exploration. Second, for the critics with double Q-network structure, the networks are trained to effectively estimate the Q-value for any specific action under a given state. Thus, the predicted Q-networks are subjected to the loss functions in terms of the Bellman residual, given as
\begin{equation} \label{eq:loss-cri}
   \mathcal{L}(\bm{\phi}_i) = \mathbb{E}_{\{\bm{s}(t), \bm{a}(t), \bm{s}(t+1)\}\sim\mathcal{D}} \left[ \left( Q_{\bm{\phi}_i}(\bm{s}(t), \bm{a}(t)) - y(t) \right)^2 \right],
\end{equation}
where the transition $\{\bm{s}(t), \bm{a}(t), \bm{s}(t+1)\}$ is extracted from the replay buffer $ \mathcal{D} $, and the target value $y(t)$ is defined as
\begin{equation}
\begin{aligned}
&	y(t) = {r}(\bm{s}(t), \bm{a}(t)) + \eta \mathbb{E}_{\bm{a}(t+1) \sim \pi(\cdot|\bm{s}(t+1))} \\
&	\left[ \min_{i} Q_{\hat{\bm{\phi}}_i}(\bm{s}(t+1), \bm{a}(t+1)) - \alpha \log \pi(\bm{a}(t+1)|\bm{s}(t+1)) \right],
\end{aligned}
\end{equation}
calculated based on the target Q-networks with $ i=1,2 $. Third, the actor network approximates the policy of the agent to determine the probability of an action for a given state. As such, the network is updated to maximize the expected Q-value while incorporating entropy regularization as
\begin{equation} \label{eq:loss-act}
\begin{aligned}
\mathcal{L}(\bm{\psi}) = \mathbb{E}_{\bm{s}(t) \sim \mathcal{D}, {\bm{a}(t)\sim\pi(\cdot|\bm{s}(t))}} \left[ \alpha \log \pi(\bm{a}(t)|\bm{s}(t)) \right. \qquad \\ 
\left. - \min_{i} Q_{\bm{\phi}_i}(\bm{s}(t), \bm{a}(t)) \right],
\end{aligned}
\end{equation}
where the exploration (via the entropy term) and exploitation (via the Q-value) are balanced for action determination.

\begin{algorithm}[t] 
\caption{SAC Learning-Based UAV Deployment} \label{alg}
\KwIn{Initial UAV position $\bm{s}_0$, SAC actor $\pi_{\bm{\psi}}$, critics $Q_{\bm{\phi}_1}$, $Q_{\bm{\phi}_2}$, target critics ${Q}_{\hat{\bm{\phi}}_1}$, ${Q}_{\hat{\bm{\phi}}_2}$, replay buffer $\mathcal{D}$, and GNN model for beamforming;}
\KwOut{Optimized UAV policy $\pi_{\bm{\psi}}$ and deployment $\bm{s}$.}
\textbf{Initialize:} $\pi_{\bm{\psi}}$, $Q_{\bm{\phi}_1}$, $Q_{\bm{\phi}_2}$, ${Q}_{\hat{\bm{\phi}}_1}$, ${Q}_{\hat{\bm{\phi}}_2}$, and $\mathcal{D}$\;
\For{each episode}{
    Reset environment and set initial state $\bm{s}_0$\;
    \For{each time step $t$}{
        Sample an action $\bm{a}(t) \sim \pi_{\bm{\psi}}(\cdot|\bm{s}(t))$\;
        Update UAV position as $\bm{s}(t+1) = \bm{s}(t) + \bm{a}(t)$\;
        Obtain the transmit beamforming vectors $\{\bm{w}_k\}_{k\in\mathcal{K}}$ using GNN model under current UAV deployment\;
        Calculate reward $r(\bm{s}(t), \bm{a}(t))$ based on the secrecy rate\;
        Store the transition $(\bm{s}(t), \bm{a}(t), r(\bm{s}(t), \bm{a}(t)), \bm{s}(t+1))$ in $\mathcal{D}$\;
        Sample a mini-batch of transitions from $\mathcal{D}$\;
        Update the predicted critic networks, actor network, and adjust entropy temperature by minimizing the loss in~(\ref{eq:loss-cri}), (\ref{eq:loss-act}), and~(\ref{eq:loss-entr}), respectively\;
        Softly update target critic networks\;
    }
}
\Return Optimized policy $\pi_{\bm{\psi}}$ and UAV deployment.
\end{algorithm}

\subsection{Training Process and Algorithm}
Based on the SAC-based learning framework, the neural networks are then trained for security-oriented UAV deployment. First, define the environment based on the communication scenario, and initialize the parameters for the established neural networks. Then, for each episode, the UAV as the agent interacts with the environment by sampling $\bm{a}(t) \sim \pi(\cdot|\bm{s}(t))$ and observing the next state $\bm{s}(t+1)$ with achieved reward ${r}(\bm{s}(t), \bm{a}(t))$. Note when evaluating the reward, the GNN-based security beamforming is exploited under current UAV deployment to calculate the system secrecy rate. The transition $(\bm{s}(t), \bm{a}(t), {r}(\bm{s}(t), \bm{a}(t)), \bm{s}(t+1))$ is obtained and stored in the replay buffer $\mathcal{D}$. With sufficiently filled replay buffer, the actor, critic, and entropy network parameters are then updated by minimizing their respective loss functions by sampling mini-batches of transitions from the replay buffer. When the predicted critic networks are periodically trained, the parameters of the corresponding target critics are updated by adopting the soft update rules. The procedures above are continued until the convergence is achieved, and the overall algorithm is summarized in Alg.~\ref{alg}. 

For algorithm implementation, we note that the GNN model for security beamforming is trained separately before the SAC reinforcement learning model training. The GNN is trained to compute optimal beamforming vectors by directly maximizing the secrecy rate across legitimate users. The adopted GNN structure features remarkable scalability, and thus can be conveniently generated to the cases under different network settings. Once the GNN is trained, it is integrated into the SAC learning process to serve as the inner-layer model. During SAC training, the GNN is used to compute the reward for each UAV position generated by the SAC agent, where the scalability of inner-layer GNN facilitates the evolution of deployment policy under SAC learning. This hierarchical training sequence not only simplifies the overall learning process but also enhances the stability and efficiency of the proposed framework.

Generally, the SAC algorithm optimizes a policy objective incorporating an entropy term, under standard assumptions, i.e., smoothness of the MDP and bounded rewards, SAC-based learning is guaranteed to converge~\cite{sac}. While in our proposal, the instantaneous reward is approximated through the GNN, which significantly complicates the rigorous analysis on the convergence from a theoretical perspective. Accordingly, we resort to numerical evidence to validate the convergence of model training, as shown later in the simulation results.
Furthermore, the complexity of the proposed deep graph reinforcement learning approach is analyzed as follows. First, the inner-layer GNN-based beamforming concerns $K$ nodes, and each node has $6N$ features. With $D$ GCN layers and $ E $ training epochs, the training complexity is $ \mathcal{O}\left( DE(K^2 + K N^2) \right) $ due to the inter-user message aggregation and the feature update operations. Accordingly, when the GNN is trained, the inference is of a complexity of $ \mathcal{O}\left( D(K^2 + K N^2) \right) $. Additionally, for the outer-layer SAC procedure, assume there are $ L $ actor/critic layers with an average $ M $ hidden units, each step involves a network complexity of $ \mathcal{O}(LM^2) $. Incorporating the GNN-based inference for reward calculation, the training complexity becomes $ \mathcal{O}( F( BD(K^2 + K N^2) + LM^2) ) $, with $ B $ being the batch size and $ F $ being the total training steps. When the overall network is trained, the inference complexity becomes $ \mathcal{O}( D(K^2 + K N^2) + LM^2) $.

\begin{figure}[!t]
   \centering
   \includegraphics[width=0.4\textwidth]{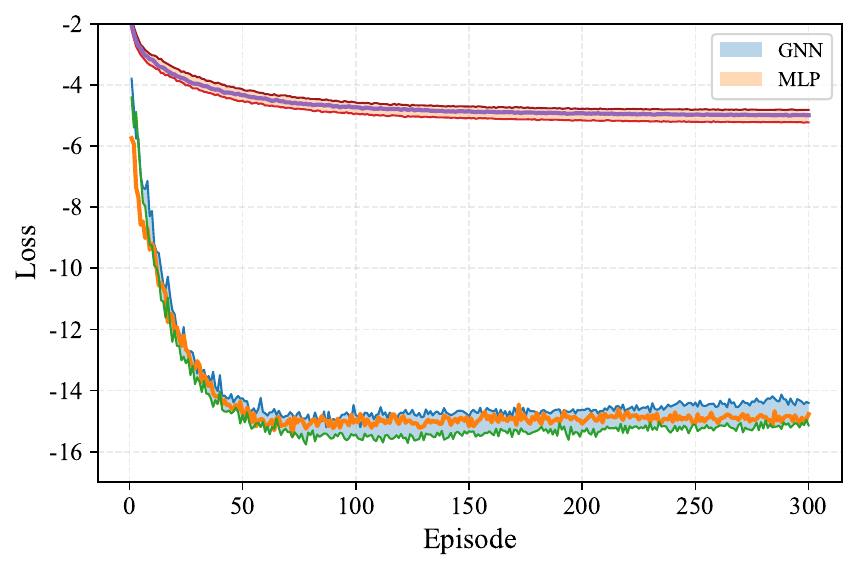} 
   \caption{Convergence of GNN model training.}
   \label{fig:conv-gnn}
\end{figure}

\begin{figure}[!t]
   \centering
   \includegraphics[width=0.4\textwidth]{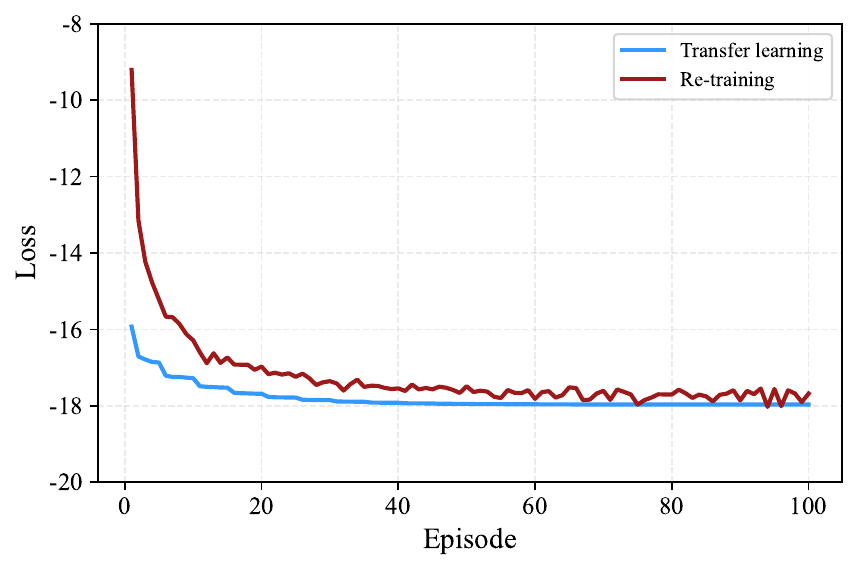} 
   \caption{Convergence of GNN model under changing scenarios.}
   \label{fig:conv-trans}
\end{figure}

\begin{figure}[!t]
   \centering
   \includegraphics[width=0.4\textwidth]{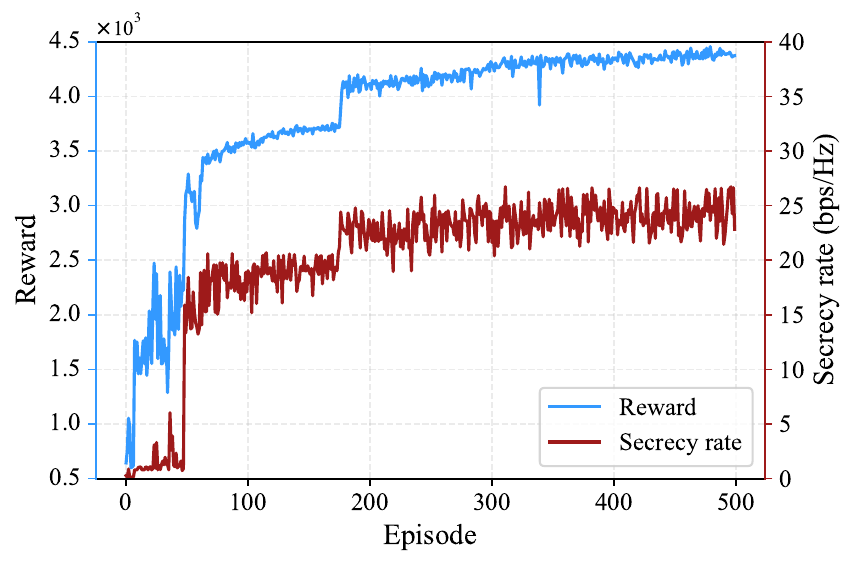} 
   \caption{Convergence of SAC model training.}
   \label{fig:conv-sac}
\end{figure}

\begin{figure}[!t]
   \centering
   \includegraphics[width=0.4\textwidth]{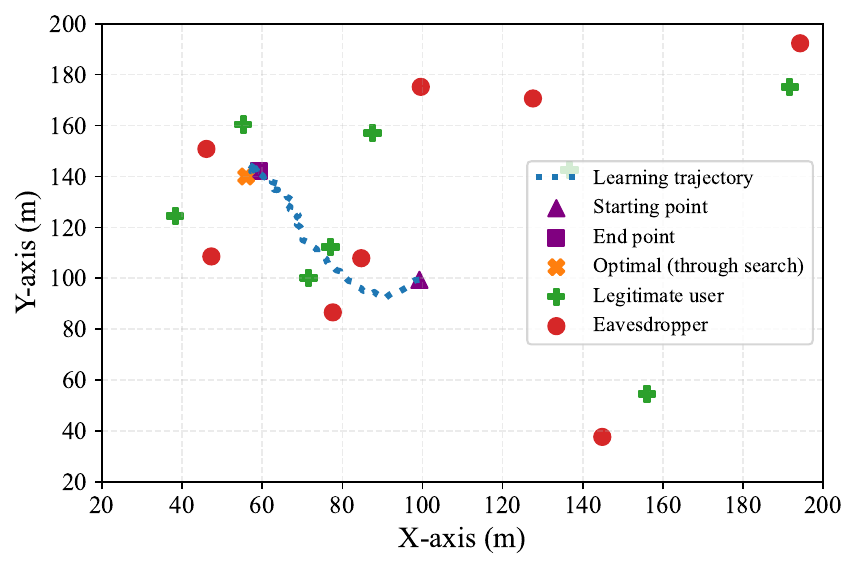} 
   \caption{Procedure to approximate the optimal deployment through SAC.}
   \label{fig:conv-traj}
\end{figure}

\begin{figure}[t]
   \centering
   \includegraphics[width=0.38\textwidth]{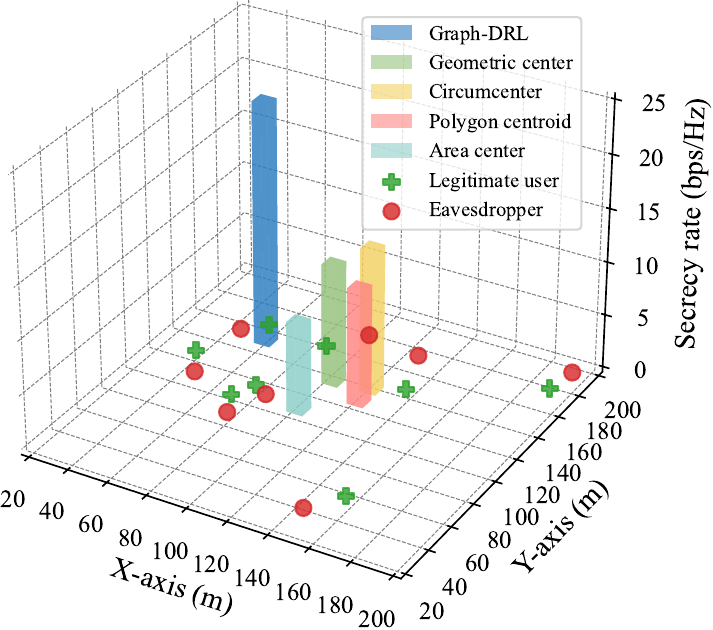} 
   \caption{Achieved secrecy rate under different deployment strategies.}
   \label{fig:deploy}
\end{figure}

\section{Simulation Results} \label{sec:sim}

\begin{table}[t] \footnotesize
    \centering
    \renewcommand{\arraystretch}{1.2}
    \setlength{\tabcolsep}{6pt}
    \caption{Simulation Parameters}
    \label{tab:sim}
    \begin{tabular}{l c | l c}
        \toprule
        \multicolumn{2}{c|}{\textbf{Network Scenario Settings}} & \multicolumn{2}{c}{\textbf{Neural Network Settings}} \\ 
        \midrule
        Area Size & 200~m$\times$200~m & GNN Learning Rate & 0.005 \\  
        No. of Users & 8 & GNN Train Episodes & 300 \\  
        No. of Antennas & 8 & GNN Batch Size & 512 \\  
        UAV Altitude & 100~m & SAC Learning Rate & 0.0003 \\  
        Rician Factor & 10~dB & SAC Train Episodes & 500 \\  
        Bandwidth & 1~Hz & SAC Mini-Batch Size & 64 \\  
        Power Budget & 1~W & SAC Buffer Size & 10\textsuperscript{6} \\  
        Noise Power & 1.2$\times$ 10\textsuperscript{-13}~W & Discount Factor & 0.99 \\  
        \bottomrule
    \end{tabular}
\end{table}

In this section, we present the simulation results to evaluate the proposed deep graph reinforcement learning approach for UAV-enabled secure communications. Specifically, we consider an area of 200~m$\times$200~m, where the users are randomly distributed with a corresponding eavesdropper located about 20~m away, and the the UAV base station is initialized at the area center with a height of 100~m. There are 8 legitimate user-eavesdropper pairs, and the UAV is equipped with 8 antennas. The air-ground UAV communication channel is considered as Rician fading with a Rician factor of 10~dB, where the line-of-sight component follows the path-loss model of 30+22$\log(d)$ in dB, with $ d $ being the distance between the transmitter and receiver~\cite{chnl}. The maximum transmit power is 1~W, and the noise power is 1.2$\times$10\textsuperscript{-13}~W. This corresponds to a typical UAV communication scenario as evaluated in may existing studies, and the settings above are used as default unless otherwise noted. Also, this is the setup to train the GNN model, which is further used to cover other scenarios without re-training. For the neural networks, we establish a GNN with 5 GCN layers. The model is trained using a learning rate of 0.005 over a maximum of 300 training episodes. For the SAC framework, the action space is defined within the interval $[\text{-1,1}]$. The SAC model is trained with a learning rate of 0.0003 and a maximum of 500 episodes, ensuring sufficient exploration and policy convergence. A summary of the main simulation settings is provided in Tab.~\ref{tab:sim}. The exiting work on GNN-based beamforming is referenced for the hyperparameter settings of the neural networks~\cite{uav-gnn-3,gnn-mark-1}, which are further adjusted with trial observations to balance the convergence speed, computation efficiency, and performance.

For performance comparison, we consider the following benchmarks. Regarding the UAV deployment, we seek for the optimal deployment through exhaustive search, and also evaluate some heuristic deployment strategies based on the node locations. Regarding the security beamforming, we consider the optimization-based scheme in~\cite{opt}, which is verified with near-optimal security performance. Also, we construct a conventional deep model with pure MLP structure to approximate the security beamforming.

\begin{figure*}[t] 
  \centerline{
  \subfigure[Performance with respect to number of users.]{
    \label{fig:num-user} 
    \includegraphics[width=5.7cm]{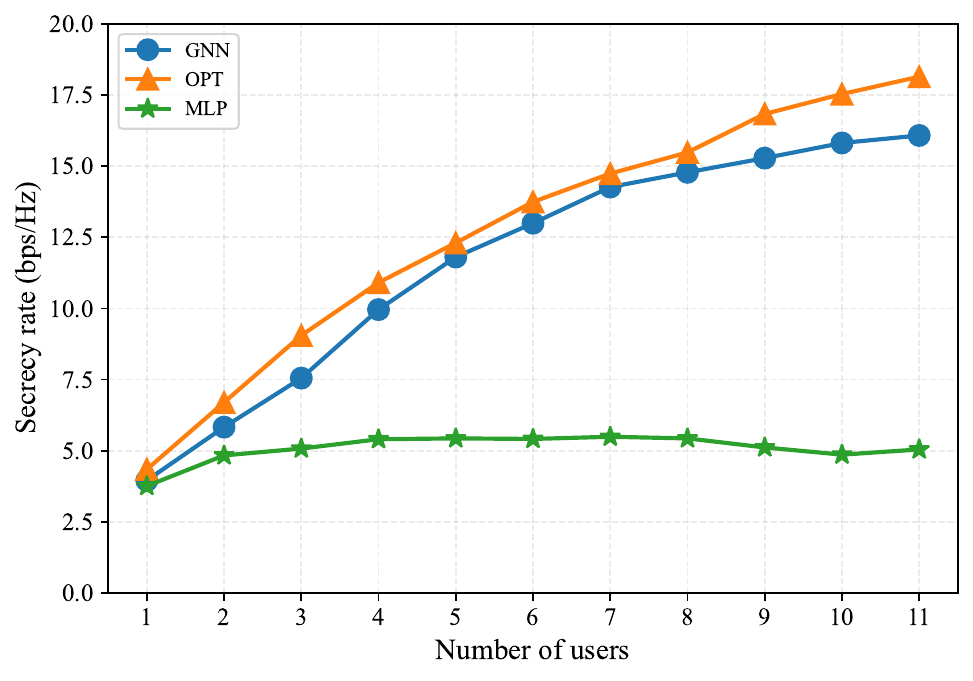}}
  \subfigure[Performance with respect to transmit power.]{
    \label{fig:pwr} 
    \includegraphics[width=5.7cm]{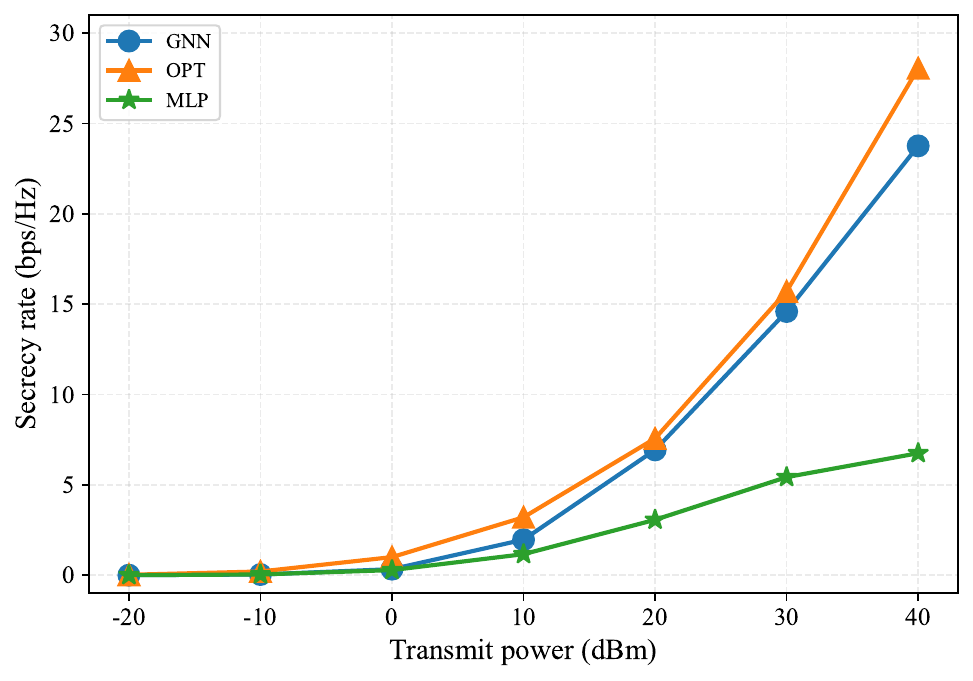}}
    \subfigure[Performance with respect to noise power.]{
    \label{fig:noise} 
    \includegraphics[width=5.7cm]{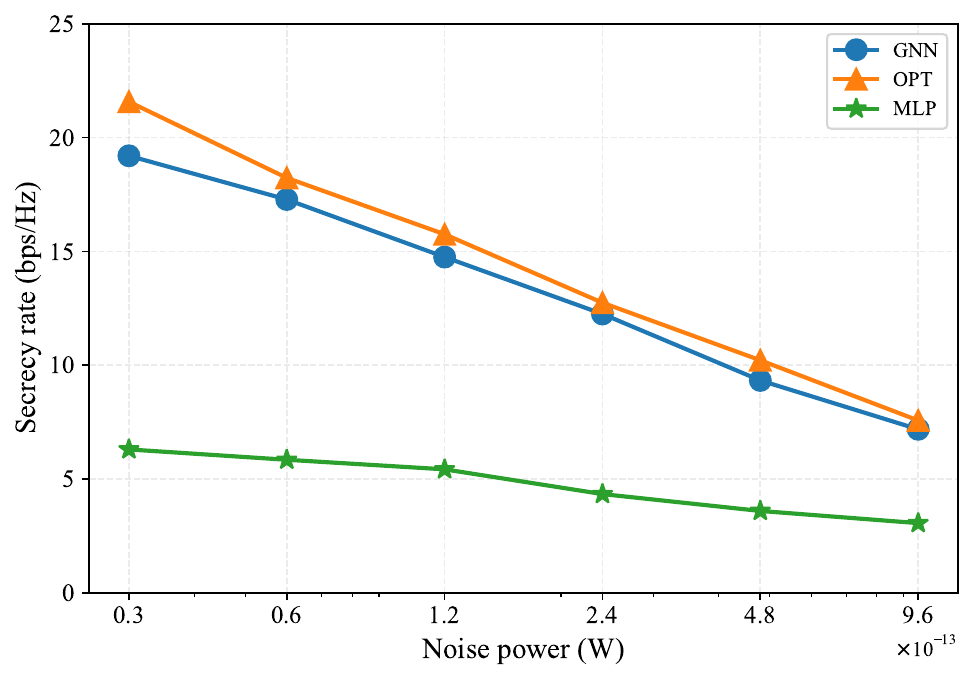}}
  }
  \caption{Performance under different network scenarios. (GNN is only trained at default setting, and generalized to cover other network configurations.)}
  \label{fig:comp} 
\end{figure*}

\subsection{Convergence}

We first show the convergence of the proposed deep graph reinforcement learning approach. Specifically, Fig.~\ref{fig:conv-gnn} depicts the training convergence of the inner-layer security beamforming model with fixed UAV deployment. We demonstrate the achieved loss functions along with the training procedure, where the lines in different groups corresponds to the trials for the GNN model (blue shaded) and MLP model (pink shaded), respectively. For both schemes, the achieved loss downgrades rapidly, indicating the effectiveness in security beamforming optimization. Furthermore, the proposed GNN model consistently outperforms the MLP baseline, achieving significantly lower loss values as training progresses. This superior performance demonstrates the capability of GNN to effectively learn graph-structured dependencies, which are essential for capturing the interference relationships among legitimate users and eavesdroppers. However, the convergence under GNN has more evident fluctuations, this is basically due to the more complicated network structure of GNN as compared with general MLP structure.

While Fig.~\ref{fig:conv-gnn} shows the case of 8 users as the default setting, we in Fig.~\ref{fig:conv-trans} extend the evaluation to the case with 12 users under GNN model. Particularly, we compare two approaches: re-training the GNN from scratch and applying transfer learning based on the pre-trained model for 8 users. As can be seen. The transfer learning approach achieves significantly faster convergence as compared with re-training a new model, and the achieved loss is also slightly lower in former scheme. In this regard, the transfer learning can be regarded as a fine tuning for the unseen scenario. This result demonstrates the effectiveness in adapting GNN to new scenarios by leveraging the prior knowledge, and highlights the practicality of transfer learning in GNN-based solutions for rapid adaptation to changing environments. Note the transfer learning is feasible thanks to the scalability of proposed GNN model, which is generally inapplicable under MLP structures.




In Fig.~\ref{fig:conv-sac}, we illustrate the convergence behavior of the SAC agent during reinforcement learning, showing the achieved reward and secrecy rate while training. Overall, both metrics improve gradually as the number of training episodes increases, indicating the adaptation of the agent for the UAV deployment task. Moreover, the correlation between reward and secrecy rate underscores the effectiveness of the former in guiding the agent to improve the security performance. Accordingly, Fig.~\ref{fig:conv-traj} illustrates the SAC learning procedure for UAV deployment for a certain trial, where the UAV starts from the area center as the initial location and gradually moves to the end position. As we also indicate the optimal deployment through exhaustive search, we can see that the learned deployment is rather close to the optimal. These results demonstrate the capability of the proposed method to approximate optimal UAV placement efficiently, and confirm the practicality of our proposal in maximizing the secrecy rate in UAV-enabled communication systems.


\subsection{Performance Comparison}

For the performance comparison, we first show the achieved secrecy rate with different deployment strategies in Fig.~\ref{fig:deploy}. Here we consider the same topology with the simulation trial in Fig.~\ref{fig:conv-traj}, and evaluate the baselines by deploying the UAV at the area center, the geometric center, the circumcenter, the polygon centroid of all the user locations. As shown in Fig.~\ref{fig:deploy}, the proposed deep graph reinforcement learning framework achieves the highest secrecy rate compared to the baselines, where the advantage can be rather significant. This results not only validate the effectiveness of the proposed framework in optimizing UAV deployment, but also highlight the importance of flexible deployment of UAV to dynamically adapt to user and eavesdropper distributions in security provisioning.

Furthermore, we evaluate the security performance under different beamforming schemes in Fig.~\ref{fig:comp}, where the proposed GNN-based beamforming is compared with the cases with optimization-based solution and MLP-based learning. Specifically, Fig.~\ref{fig:num-user} illustrates the secrecy rate with respect to the number of users in the network. Here, we emphasize that the proposed GNN model is trained for the case with 8 users, while the results for other scenarios are directly generalized based on the trained model without retraining. As can been seen, the proposed GNN-based framework approaches the performance of the optimization-based method, while achieving significantly higher secrecy rates compared to the MLP baseline across all tested scenarios. Also, as the number of users increases, the gap between the GNN model and the optimization-based approach becomes more pronounced, which reveals the slightly downgraded representation capability of GNN in more complicated cases with heavier interference. In contrast, the MLP model struggles to model complex user interactions effectively in larger networks, and thus the overall performance is rather limited. The results reveal the generalization capability of the proposed graph-based learning scheme, which directly covers the new network scenarios without retraining or additional computation overhead. More importantly, the near-optimal performance can be also preserved in new cases, indicating that the graph structure and mutual interference among the users have been effectively learned and represented in the trained networks.

Fig.~\ref{fig:pwr} shows the secrecy rate versus transmit power, where we train the GNN model with a transmit power of 1~W, and generalized the model to other cases. Meanwhile, Fig.~\ref{fig:noise} presents the secrecy rate with respect to the noise power, where we train the GNN model with noise power of 1.2$\times$10\textsuperscript{-13}~W and scale to cover other cases. The results in these two figures show the similar trend that, our proposal closely approaches the optimization-based solution, while substantially dominating the MLP baseline. Also, naturally, with higher transmit power or lower background noise, the achieved secrecy rate can be higher under all schemes. The results indicate that for interference-limited scenarios, the inherent ability to represent the interfering relationship is critical for the neural network to interpret effective beamforming strategy. Moreover, we can conveniently generalized a trained GNN model to diverse scenarios, including variations in network size, transmit power, and noise power, with maintained near-optimal secrecy performance, verifying the scalability, adaptability, and robustness of the GNN-based framework.

\begin{figure}[t]
   \centering
   \includegraphics[width=0.42\textwidth]{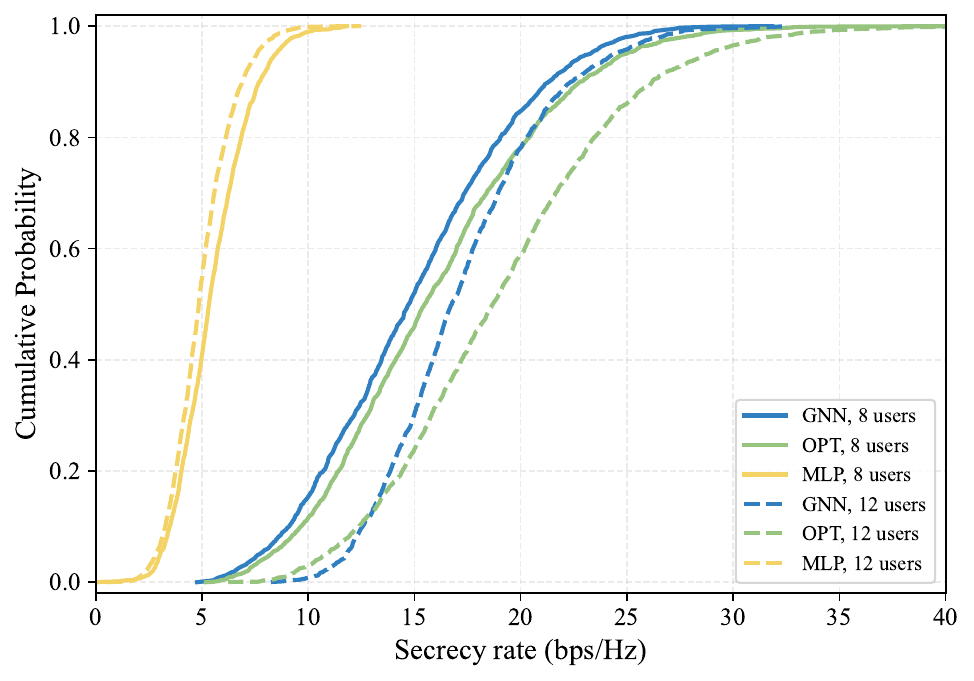} 
   \caption{CDF of achieved rate for different numbers of users.}
   \label{fig:dist}
\end{figure}

In Fig.~\ref{fig:dist}, we presents the cumulative distribution function (CDF) of the achieved secrecy rates for two scenarios: a trained GNN model with 8 users and a generalized GNN model applied to 12 users, which are also compared with the optimization-based method and the MLP baseline. For the 8-user case, the achieved distribution of secrecy rate under GNN scheme closely matches that of the optimization approach, significantly outperforming the MLP baseline, which is in consistence with the results presented before. In the 12-user case, where the GNN model is generalized without retraining, the CDF curve shows that the GNN continues to perform robustly while achieving near-optimal secrecy rates. In particular, we can see that the users with relatively higher and relatively lower rate under generalized GNN model are less than those under optimization strategy, while the number of users with moderate secrecy rate is higher. This indicates that the generalized GNN model achieves a relatively more ``averaged'' performance across all users, and thus induces a slightly higher performance gap as compared with the optimum. Yet the gap is still relatively small, and thus the results validate the scalability and adaptability of the GNN-based framework.

\begin{figure}[t]
   \centering
   \includegraphics[width=0.42\textwidth]{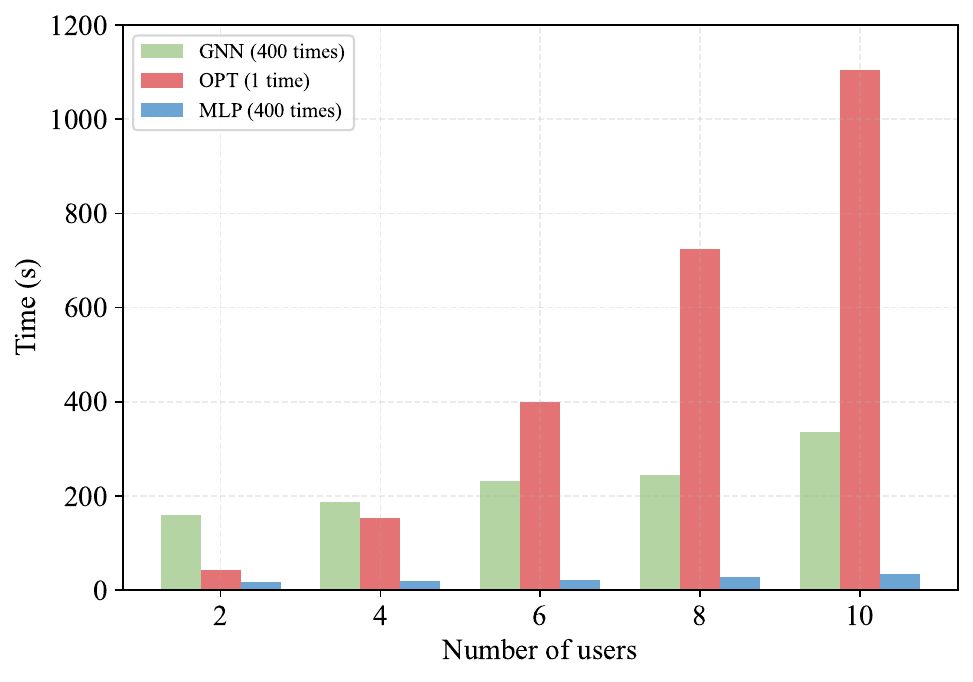} 
   \caption{Time consumption for different approaches.}
   \label{fig:time}
\end{figure}

Fig.~\ref{fig:time} compares the computational time required by the proposed GNN-based framework, the optimization-based method, and the MLP model. As the neural network model execution can be finished rather quickly, we show the time taken for 400 runs. As expected, the GNN framework demonstrates significantly lower time consumption compared to the optimization-based method, even when executed 400 times, highlighting its computational efficiency. Moreover, as the number of users increases, the time required for the GNN model remains relatively stable, demonstrating the scalability to larger network sizes. In contrast, the time required for the optimization grows rather evidently due to its inherent computational complexity as a function of number of users. The MLP baseline also shows low time consumption but delivers inferior performance compared to the GNN, as indicated in previous results. The result confirms the practicality of the GNN-based framework for real-time applications, where efficient computation with the scalability to different numbers of users corroborate our proposal as a robust solution for UAV-enabled secure multi-user communications.

As a brief summary, we can see that the proposed deep graph reinforcement learning-based approach achieves near-optimal security performance in various network scenarios, while significantly reducing the implementation complexity to support real-time applications. More importantly, the graph-based learning mechanism enables compelling scalability, allowing the capability to handle large and dynamic environments effectively.

\section{Conclusion} \label{sec:con}

In this work, we have proposed a deep graph reinforcement learning framework to address the physical layer security issue for UAV-enabled multi-user communication. By decomposing the problem into two layers, we have designed the GNN-based security beamforming in the inner layer and SAC-based UAV deployment in the out layer to achieve security enhancement. The results have demonstrated the effectiveness of the proposed learning framework, which achieves near-optimal secrecy performance across varying network environments with significantly higher execution efficiency. Moreover, the framework can be effectively generalized to unseen scenarios without requiring substantial retraining. These results have collectively established the proposed framework as a robust and scalable solution for secure communication in dynamic UAV-enabled multi-user networks.

\bibliographystyle{IEEEtran}
\bibliography{main}

\end{document}